\documentclass[hyper]{JHEP} 

\usepackage{epsfig}

\newcommand\fverb{\setbox\pippobox=\hbox\bgroup\verb}
\newcommand\fverbdo{\egroup\medskip\noindent%
			\fbox{\unhbox\pippobox}\ }
\newcommand\fverbit{\egroup\item[\fbox{\unhbox\pippobox}]}
\newbox\pippobox
\title{Non-BPS Dp-Brane in the Background of
NS5-Branes on Transverse $R^3\times S^1$}

\author{by J. Kluso\v{n}\\
	 Department of Theoretical Physics and Astrophysics\\
                   Faculty of Science, Masaryk University\\
Kotl\'{a}\v{r}sk\'{a} 2, 611 37, Brno\\
Czech Republic\\
	E-mail: \email{klu@physics.muni.cz}}

\preprint{\hepth{0411014}}

\abstract{This paper is devoted to the study of
non-BPS Dp-branes in the presence of 
NS5-branes on the transverse $R^3\times S^1$.
We will formulate the tachyon effective action
in this background and then we will
discuss  its properties.
 Then we will
study the solutions of the equations of motion that
describe lower dimensional BPS and non-BPS
D-branes.}


\def\mT{\mathcal{T}}
\def\bM{\mathbf{M}}
\def\bZ{\mathbf{Z}}
\def\bz{\mathbf{z}}
\def\bA{\mathbf{A}}
\def\mV{\mathcal{V}}
\def\bI{\mathbf{I}}
\def\bT{\mathbf{T}}
\begin{document}
\section{Introduction}\label{first}
It is well known that type IIA (IIB) string
theories contain two types of D-branes:
the BPS Dp-branes, which have
even (odd)p in type IIA (IIB) theories
and unstable, non-BPS Dp-branes with
odd (even)p in the type IIA (IIB) case,
for review see \cite{Sen:2004nf,Sen:1999mg}. 
Non-BPS
D-branes are very important in string
theory. For example,  the BPS
D-branes can be thought as solitons
in the worldvolume theory of non-BPS ones
\cite{Witten:1998cd,Witten:2000cn,Horava:1998jy}.
However, as was stressed recently in
\cite{Kutasov:2004ct} there are many open questions
about them that remain unanswered. 
For example, it is very remarkable that
many aspects of the tachyon dynamics 
can be captured by a spacetime
effective action of Dirac-Born-Infeld (DBI) type
\cite{Sen:1999md,Garousi:2000tr,Bergshoeff:2000dq,
Kluson:2000iy,Lambert:2003zr,
Kutasov:2003er,Niarchos:2004rw}
\begin{equation}
S=-T^{non}_p\int d^{p+1}\xi
\frac{1}{\cosh\frac{T}{\sqrt{2}}}
\sqrt{-\det G} \ ,
\end{equation}
where $T^{non}_p$ is tension of non-BPS Dp-brane
and $G$ is induced metric on the brane
\begin{equation}\label{inmet}
G_{\mu\nu}=\eta_{\mu\nu}+\partial_{\mu}T
\partial_{\nu}T+\partial_{\mu}Y^I
\partial_{\nu}Y^I \ .
\end{equation}
The scalar fields $Y^I \ (I=p+1,\dots,9)$
living on the worldvolume of the brane parameterise
its location in the transverse space. The form
of the induced metric (\ref{inmet}) suggests that
the tachyon direction in field space should
be treaded as an extra dimension of space,
like $Y^I$, however then there is an important
question: what is the meaning of the tachyon
potential $V(T)$? 

In two recent papers by D. Kutasov
\cite{Kutasov:2004dj,Kutasov:2004ct}
\footnote{Similar problems were
discussed in \cite{Kluson:2004xc,Saremi:2004yd,Sahakyan:2004cq,
Ghodsi:2004wn,Panigrahi:2004qr,Yavartanoo:2004wb}.}
the precise
analogy  
between the BPS D-brane moving
in the background of NS5-branes and the
tachyon dynamics on non-BPS Dp-brane
was demonstrated. In particular,
 in \cite{Kutasov:2004ct}
the dynamics of BPS D-brane propagating
in the near horizon limit of  $NS5$-branes with
the transverse space $R^3\times S^1$ was
considered as an useful toy model of the 
non-BPS D-brane. It was shown that
from the point of view of an observer
living on the  5+1 dimensional worldvolume
of fivebranes the BPS D-branes in the full
theory give rise to two kinds of the objects
in five dimensions. One consists D-branes
whose worldvolume lies entirely inside
the worldvolume of fivebranes. These 
D-branes are non-BPS-while both fivebranes
and the BPS D-branes separately preserve 
16 supersymmetries, a background that contains
both of them breaks all supersymmetry. 
The second kind consists D-branes that wrap
the circle transverse to the fivebranes.  These
D-branes preserve eight supercharges and are
BPS. In summary, two kinds of D-branes emerge
from different orientations of BPS D-branes
in the space transverse to fivebranes.

In this paper we will continue  the study of
this interesting problem. Namely, we will consider
a non-BPS D-brane embedded in the background
of $k$ $NS5$-branes on transverse $R^3\times S^1$.
 We will explicitly show that 
 the  
tachyon effective  action in this   background
is invariant under special
transformation that maps the   tachyon
mode $T$-that is presented on the 
worldvolume of a non-BPS Dp-brane even in the 
flat spacetime-to 
the new tachyon field $\mT$-that arises 
from the field redefinition of the worldvolume field
$y$ that parameterises the position of 
a non-BPS Dp-brane on the circle $S^1$. 
 The existence of this symmetry,
even if its physical origin is unclear to us,
really suggests that it is correct
to consider the tachyon mode
as an additional embedding coordinate.
However  the origin of this possible additional
dimension is unclear at present.
 
After the discussion of the general properties
of the tachyon effective action in the fivebranes
background 
   we will study some solutions of its equations
of motion. We will show that
these solutions  describe both non-BPS and
BPS lower dimensional D-branes that
are embedded  in the fivebrane
background with the transverse space
$R^3\times S^1$. These solutions explicitly
demonstrate that all branes in the fivebranes
background arise through the tachyon condensation
on the worldvolume of a non-BPS D-brane in
the same way as D-branes in the flat spacetime
can be thought as solitonic solutions on the
higher dimensional  non-BPS D-brane. 

This paper is organised as follows. In the
next section (\ref{second}) we will 
study the properties of a non-BPS Dp-brane
in the NS5-branes background with transverse
space $R^3\times S^1$. Then in section
(\ref{third}) we will analyse some solutions
of the equation of motion and we will give
their physical interpretation.
Finally, in conclusion (\ref{fourth}) we will outline
our results and suggest possible extension of
this work. 
\section{Tachyon effective action in 
the presence of
NS5-branes on transverse $R^3\times S^1$}
\label{second}
To begin with we give a brief
description of 
the system of $k$ $NS5$-branes on transverse 
$R^3\times S^1$ which we will
label with coordinates $(\bZ,Y)$ 
with $\bZ=(Z^1,Z^2,Z^3)\in
R^3$ and $Y\sim Y+2\pi R$ 
where $R$ is radius $S^1$. The
fivebranes are located at
 points $\bZ=Y=0$. The background
around them is \cite{Polchinski:1998rr}
\begin{eqnarray}\label{BS}
ds^2=dx^{\mu}dx_{\mu}+H(\bZ,Y)
(d\bZ^2+dY^2) \ , \nonumber \\
e^{2(\Phi-\Phi_0)}=H(\bZ,Y)
\ , \nonumber \\
\end{eqnarray}
where $x^{\mu}\in R^{5,1}$ label 
the worldvolume
of fivebranes and where $\Phi_0$ is related 
to the string
coupling constant $g_s$ as $g_s=\exp \Phi_0$. The
harmonic function $H$ in (\ref{BS}) has the form
\begin{equation}
H=1+k\sum_{n=-\infty}^{\infty}
\frac{1}{(Y-2\pi Rn)^2+\bZ^2}\ . 
\end{equation}
We will be interested in the study of the system in
the near-horizon limit that can be defined by rescaling
all  distances by the factor $g_s$ 
\begin{equation}
\bZ=g_s\bz \ , R=g_sr \ , Y=g_s y
\end{equation}
and sending $g_s\rightarrow 0$ while keeping the
rescaled distances $(\bz,y,r)$ fixed. This leads to the
background
\begin{equation}\label{nhg}
ds^2=dx^{\mu}dx_{\mu}+h(\bz,y)(d\bz^2+dy^2) \ ,
e^{2\Phi}=h(\bz,y) 
\end{equation}
with
\begin{equation}\label{nhh}
h(\bz,y)=k\sum_{n=-\infty}^{\infty}
\frac{1}{(y-2\pi rn)^2+\bz^2}=
\frac{k}{2r|\bz|}
\frac{\cosh\left(\frac{|\bz|}{r}\right)}
{\cosh\left(\frac{|\bz|}{r}\right)
-\cos\left(\frac{y}{r}\right)} \ . 
\end{equation}
We now place non-BPS Dp-brane whose worldvolume
is embedded entirely in $R^{5,1}$ in the geometry
(\ref{nhg}) and (\ref{nhh}). The action for a
non-BPS Dp-brane
in this background takes the form
\begin{eqnarray}\label{nonact2}
S=-\int d^{p+1}\xi \frac{V(T)}{\sqrt{h(\bz,y)}}
\sqrt{-\det G_{\mu\nu}}= \ \nonumber \\
=-\int d^{p+1}\xi \sqrt{-\det \eta}
\frac{V(T)}{\sqrt{h(\bz,y)}}
\sqrt{\det(\bI+\bM)} \ , \nonumber \\
\end{eqnarray}
where 
\begin{equation}
G_{\mu\nu}=\eta_{\mu\nu}+
h(\bz,y)\left(\partial_{\mu}z^i
\partial_{\nu}z^i+\partial_{\mu}y\partial_{\nu}y
\right)
+\partial_{\mu}T\partial_{\nu}T 
\end{equation}
and where we have also  introduced 
$(n+1)\times (n+1)$ unit matrix 
$\bI^{\mu}_{\nu}=\delta^{\mu}_{\nu}$
together with 
$(n+1)\times (n+1)$ matrix $\bM$ 
\begin{equation}
\bM^{\mu}_{\nu}=h(\bz,y)(\partial^{\mu}z^i
\partial_{\nu}z^i+\partial^{\mu}y\partial_{\nu}y)
+\partial^{\mu}T\partial_{\nu}T \ . 
\end{equation}
The action  (\ref{nonact2}) describes the 
non-BPS Dp-brane that
is localised in the transverse space labelled with
$\bz,y$. As in the case of BPS Dp-brane
studied in \cite{Kutasov:2004ct}
we will be interested in the study of the
dynamics of the mode $y$. 
For that reason
we should show that $\bz$ can be put 
in the values that solve their 
 equations of motions that 
arise from
 (\ref{nonact2})
\begin{eqnarray}\label{eqSx}
-\frac{1}{2h^{3/2}}\partial_{z^i}h V(T)
\sqrt{\det(\bI+\bM)}-\nonumber \\
-\partial_{\kappa}\left[\eta^{\kappa\mu}
\sqrt{h}V(T)\partial_{\nu}z^i(\bI+\bM)^{-1\nu}_{\mu}\sqrt{\det
(\bI+\bM)}\right]+\nonumber \\
+\frac{V(T)}{\sqrt{h}}\partial^{\mu}x^m\partial_{\nu}x^m
\partial_{z^i}h (\bI+\bM)^{-1\nu}_{\mu}
\sqrt{\det(\bI+\bM)}=0 \ , \nonumber \\
\end{eqnarray}
where $x^m\equiv(\bz,y)$.
For constant
$\bz$ the equation of motion 
(\ref{eqSx}) takes simple form
\begin{equation}
\partial_{z^i} h=0 \ \ . 
\end{equation}
Using the form of 
$h$ given in (\ref{nhh})
it is easy to see that this equation
has the solution $z^i=0$.
Then  one can  place 
the fields $z^i$ in
their minimum  at $z^i=0$ and   
consider the 
dynamics of $y$ and $T$ only. 
Then the
non-BPS Dp-brane action 
takes the form
\begin{equation}\label{nonact3}
S=-\int d^{p+1}\xi 
\frac{V(T)}{\sqrt{h(y)}}\sqrt{-\det(\eta_{\mu\nu}
+h\partial_{\mu}y\partial_{\nu}y+
\partial_{\mu}T\partial_{\nu}T)} \ ,
\end{equation}
where $V(T)$ is equal to
\begin{equation}
V(T)=\frac{T^{non}_p}{\cosh\frac{T}{\sqrt{2}}} \ ,
\end{equation}
and where $T^{non}_p$ is defined such
that the tension of a non-BPS Dp-brane
at flat spacetime is $T^{non}_p/g_s$. Note
that $T^{non}_p$ is related to the quantity
that appears in the action for BPS Dp-brane as
$T^{non}_p=\sqrt{2}T^{BPS}_p$. 

Now, following \cite{Kutasov:2004dj,Kutasov:2004ct}
we introduce   
"new" tachyon field $\mT$ that is related to
$y$ through the relation 
\begin{equation}
\frac{d\mT}{dy}=
\sqrt{h(y)}=\frac{\sqrt{k}}
{2r\sin\frac{y}{2r}} \  .
\end{equation}
This differential equation has the
 solution
\begin{equation}
e^{-\frac{\mT}{\sqrt{k}}}=
\frac{\cos\frac{y}{4r}}{\sin\frac{y}{4r}}+C_0 \ .
\end{equation}
Now if we demand that 
 for $y=\pi r$  the tachyon field $\mT$
is equal to zero we get   $C_0=0$. 
Then we obtain
\begin{equation}
h(y(\mT))=\frac{1}{4r^2\cosh 
\frac{\mT}{\sqrt{k}}} \
\end{equation}
and hence  the tachyon effective action
(\ref{nonact3})  can be written as
\begin{eqnarray}\label{mTT}
S=-\int d^{p+1}\xi 
\mathcal{V}(T,\mT)\sqrt{-\det(\eta_{\mu\nu}
+\partial_{\mu}\mT\partial_{\nu}\mT+
\partial_{\mu}T\partial_{\nu}T)}= 
\nonumber \\
=-\int d^{p+1}\xi
\mathcal{V}(T,\mT)
\sqrt{\det(\bI+
\bM)} \ , \nonumber \\
\mathcal{V}(\mT,T)=\frac{\tau_p}{
\cosh\frac{\mT}{\sqrt{k}}\cosh
\frac{T}{\sqrt{2}}} \ , \nonumber \\
(\bI+\bM)_{\mu}^{\nu}
=\delta^{\nu}_{\mu}+
\eta^{\mu\kappa}
\partial_{\kappa}
\mT\partial_{\nu}\mT+
\eta^{\mu\kappa}\partial_{\kappa}
T\partial_{\nu}T \ ,
\nonumber \\
\end{eqnarray}
where
\begin{equation}
\tau_p=\frac{2T_p^{non}R}
{\sqrt{k}g_s} \ .
\end{equation}
Before we proceed to the solution
of the equation of motions that arise
from (\ref{mTT}) we would like
to say few words about
 symmetries of the
action (\ref{mTT})
 that relate $\mT$ with $T$. In
particular, we see that this action
is invariant under  the transformation
that maps $T\ , \mT , A\equiv k \ ,
B\equiv 2 $ into the
new fields $T'  , \mT'$ and new parameters
$A',B'$ given as 
\begin{equation}
\mT'=T \ , T'=\mT \ ,
A'=B \ , B'=A \ 
\end{equation}
so that $S(T',\mT',A',B')=S(T,\mT,A,B)$. 
In fact, one can find more general
symmetry of the action (\ref{mTT}).
For that reason let us 
introduce two-component vectors $X^I,Y^I \ ,
I=1,2$ defined as
\begin{equation}
X=\left(\frac{1}{\sqrt{k}},\frac{1}{\sqrt{2}}\right) \  ,
Y=\left(\frac{1}{\sqrt{k}},-\frac{1}{\sqrt{2}}\right) 
\end{equation}
that allow us to rewrite the tachyon potential
$\mV(T,\mT)$ as 
\begin{eqnarray}
\frac{\tau_p}{
\cosh\frac{\mT}{\sqrt{k}}
\cosh\frac{T}{\sqrt{2}}}=
\frac{2\tau_p}{
\left(\cosh(\frac{\mT}{\sqrt{k}}+
\frac{T}{\sqrt{2}})+
\cosh(\frac{\mT}{\sqrt{k}}-
\frac{T}{\sqrt{2}})\right)}=
\nonumber \\
=\frac{2\tau_p}{
\left(\cosh(X^I\delta_{IJ}\bT^J)+
\cosh(Y^I\delta_{IJ}\bT^J)
\right)} \ , 
\nonumber \\
\end{eqnarray}
where we have introduced two-component
vector $\bT=(\mT,T)$. In this notation the
tachyon effective action (\ref{nonact3})
takes the form
\begin{equation}\label{tachS}
S=-\int d^{p+1}\xi  \frac{2\tau_p}
{\left(\cosh(X^I\delta_{IJ}\bT^J)+
\cosh(Y^I\delta_{IJ}\bT^J)
\right)}\sqrt{-\det(\eta_{\mu\nu}
+\partial_{\mu}\bT^I\partial_{\nu}
\bT^J\delta_{IJ})} \ . 
\end{equation}
Now 
it is easy to see that this action is invariant
under following transformations 
\begin{equation}\label{sym}
X'^I=\Lambda^I_JX^J \ ,
Y'^I=\Lambda^I_JY^J \ , \bT'^I=
\Lambda^I_J\bT^J \ , 
\end{equation}
where $\Lambda^I_J$ obeys 
$\Lambda^I_K\delta_{IJ}\Lambda^J_L=
\delta_{KL}$.

We mean  that an existence of the
 transformation (\ref{sym}) is very 
attractive since it relates the
tachyon field $T$ with the field
$\mT$ and then in some sense
suggests the geometrical nature of $T$.
  Unfortunately it
 is also  clear that
this form of 
 symmetry is rather unusual and
its possible physical origin is unclear
at present. In fact we do not understand
how $T$ and $\mT$ could be related
in such a simple way when we know that
their origin is completely different. 
Secondly, we also do not understand the
physical meaning of 
the vectors $X$ and
$Y$ defined above. 
On the other hand one can hope that discovery
of all possible symmetries of the 
tachyon effective action could be
helpful for better understanding of the meaning
of the tachyon in string theory.
 
\section{Solutions of the equations of motion}
\label{third}
In this section we would like to study the equations
of motion for $T$ and $\mT$ 
that arise from the action (\ref{mTT})
\begin{eqnarray}\label{eqmTT}
-\frac{\sinh\frac{\mT}{\sqrt{k}}}
{\sqrt{k}\cosh^2\frac{\mT}{\sqrt{k}}
\cosh\frac{T}{\sqrt{2}}}\sqrt{\det
(\bI+\bM)}-
\partial_{\mu}\left[
\frac{1}{\cosh\frac{\mT}{\sqrt{k}}
\cosh\frac{T}{\sqrt{2}}}\frac{\eta^{\mu\kappa}
\partial_{\nu}\mT(\bI+\bM)^{-1 \nu}_{\kappa}}
{\sqrt{\det(\bI+\bM)}}\right]=0 \ , \nonumber \\
-\frac{\sinh\frac{T}{\sqrt{2}}}
{\sqrt{2}\cosh\frac{\mT}{\sqrt{k}}
\cosh^2\frac{T}{\sqrt{2}}}\sqrt{\det
(\bI+\bM)}-
\partial_{\mu}\left[
\frac{1
}{\cosh\frac{\mT}{\sqrt{k}}
\cosh\frac{T}{\sqrt{2}}}
\frac{\eta^{\mu\kappa}
\partial_{\nu}T(\bI+\bM)^{-1  \nu}
_{\kappa}}
{\sqrt{\det(\bI+\bM)}}\right]=0 
\ . \nonumber \\
\end{eqnarray}
For next purposes it will be also 
useful to  calculate the
stress energy tensor 
from the action (\ref{mTT}). 
In order
to do this we replace the flat 
worldvolume metric
$\eta_{\mu\nu}$ with arbitrary metric $g_{\mu\nu}$.
Using now the fact that for the action of the form
\begin{equation}
S=-\int d^{p+1}\xi \sqrt{-g}\mathcal{L}
\end{equation}
the stress energy tensor 
$T_{\mu\nu}=-\frac{2}{\sqrt{-g}}\frac{\delta S}{\delta g^{\mu\nu}}$
is equal to
\begin{equation}
T_{\mu\nu}=-g_{\mu\nu}\mathcal{L}+
2\frac{\delta \mathcal{L}}{\delta g^{\mu\nu}} 
\end{equation}
we obtain from (\ref{mTT}) 
\begin{equation}\label{strenergy}
T_{\mu\nu}=-\eta_{\mu\nu}\mathcal{V}
\sqrt{\det(\bI+\bM)}+\mathcal{V}(\partial_{\nu}\mT
\partial_{\kappa}\mT+\partial_{\nu}T
\partial_{\kappa}T)(\bI+\bM)^{-1\kappa}_{\mu}
\sqrt{\det(\bI+\bM)}  \ . 
\end{equation}
Now we are ready to study some  solutions
of the equation of motions (\ref{eqmTT}). 
We begin with the case when 
 $\mT$ is time dependent while the tachyon $T$
depends on one spatial coordinate $x\equiv \xi^1$. Then 
\begin{equation}\label{Mts}
\bM=\left(\begin{array}{ccc}
-\dot{\mT}^2 & 0 & 0 \\
0 &T'^2 & 0 \\ 
0 & 0 & 0 \\ \end{array}\right) \ ,
\det(\bI+\bM)=(1-\dot{\mT}^2)(1+T'^2) \ , 
T'\equiv \frac{dT}{dx} \   
\end{equation}
and hence components of the stress energy tensor
are equal to
\begin{eqnarray}
T_{00}=\frac{\mathcal{V}(1+T'^2)}
{\sqrt{(1-\dot{\mT}^2)(1+T'^2)}} \ , T_{0i}=0 \ ,
i=1,\dots, p\nonumber \\ 
T_{xx}=-\frac{\mathcal{V}(1-\dot{\mT}^2)}
{\sqrt{(1-\dot{\mT}^2)(1+T'^2)}} \ ,
T_{ix}=T_{xi}=0 \ , i=2,\dots, p \nonumber \\
T_{ij}=-\delta_{ij}\mathcal{V}
\sqrt{(1-\dot{\mT}^2)(1+T'^2)} \ , 
i , j=2,\dots, p \ .  \nonumber \\
\end{eqnarray}
Now for the matrix
$\bM$ given in (\ref{Mts}) 
 the equations of motion (\ref{eqmTT})
  take the form
\begin{eqnarray}\label{eqmTT1}
\frac{\sqrt{1+T'^2}}{\cosh
\frac{T}{\sqrt{2}}}
\left[-\frac{\sinh\frac{\mT}{\sqrt{k}}
\sqrt{1-\dot{\mT}^2}}
{\sqrt{k}\cosh^2\frac{\mT}{\sqrt{k}}}
+\partial_0\left(
\frac{1}{\cosh\frac{\mT}{\sqrt{k}}}
\frac{\partial_0\mT}
{\sqrt{1-\dot{\mT}^2}}\right)
\right]=0
\nonumber \\
\frac{\sqrt{1-\dot{\mT}^2}}{\cosh
\frac{\mT}{\sqrt{k}}}
\left[-\frac{\sinh\frac{T}{\sqrt{2}}
\sqrt{1+T'^2}}
{\sqrt{2}\cosh^2\frac{T}{\sqrt{2}}}
-\partial_x\left(
\frac{1}{\cosh\frac{T}{\sqrt{2}}}
\frac{\partial_xT}
{\sqrt{1+T'^2}}\right)
\right]=0 \ . \nonumber \\
\end{eqnarray}
These expressions explicitly show that
$\mT$ and $T$ decouple.  In particular,
let us consider
 the last equation in (\ref{eqmTT1}). If we
define 
\begin{equation}\label{Vx}
V(T,x)=\frac{1}{\cosh\frac{T}
{\sqrt{x}}}
\end{equation}
 then
 the expression in the bracket can be written as
\begin{eqnarray}
\frac{\delta V(T,2)}{\delta T}\sqrt{1+T'^2}
-\partial_x\left(\frac{V(T,2)\partial_xT}
{\sqrt{1+T'^2}}\right)=0
\Rightarrow \nonumber \\
\Rightarrow 
\partial_x\left(\frac{V(T,2)}{\sqrt{1+T'^2}}\right)=0
\Rightarrow 
\frac{V(T,2)}{\sqrt{1+T'^2}}=p \ ,  \nonumber \\
\end{eqnarray}
where $p$ is an integration constant. 
Using now the specific form of the
potential $V(T,2)$ given in (\ref{Vx})
 it is easy to find the dependence
of $T$ on $x$
\begin{equation}
\sinh \frac{T}{\sqrt{2}}=
\frac{\sqrt{1-p^2}}{p}\sin 
\frac{x}{\sqrt{2}} \ .
\end{equation}
For such a configuration 
the spatial 
dependent energy density is
equal to
\begin{eqnarray}
\rho(x)=T_{00}(x)=\frac{\tau_p
V(\mT,k)}{\sqrt{1-\dot{\mT}^2}}
V(T,2)\sqrt{1+T'^2}
=\nonumber \\
=\frac{\tau_pV(\mT,k)}
{\sqrt{1-\dot{\mT}^2}}
\frac{p}{p^2+(1-p)^2\sin^2 
\frac{x}{\sqrt{2}}} \ .\nonumber \\
\end{eqnarray}
The physical interpretation of this
solution is in terms of an
array of D(p-1)-branes and 
D(p-1)-antibranes 
\cite{Kim:2003ma,Brax:2003rs,Kim:2003in}
 that move toward
to the 
worldvolume of $NS5$-branes. 
To see this more explicitly 
let us now solve the equation 
of motion for
$\mT$ (\ref{eqmTT1}) 
that can be written
as
\begin{eqnarray}
-\frac{\delta V(\mT,k)}{\delta \mT}
\sqrt{1-\dot{\mT}^2}+
\partial_0
\left(\frac{V(\mT,k)\dot{\mT}}
{\sqrt{1-\dot{\mT}^2}}\right)=0
\Rightarrow
\nonumber \\
\Rightarrow
\partial_0\left(\frac{V(\mT,k)}{\sqrt{1-\dot{\mT}^2}}
\right)=0
 \Rightarrow
 \frac{V(\mT,k)}{\sqrt{1-\dot{\mT}^2}}=e \ .
 \nonumber \\
\end{eqnarray}
Using again (\ref{Vx}) we obtain
\begin{eqnarray}
\sinh\frac{\mT}
{\sqrt{k}}=
\frac{\sqrt{e^2-1}}{e}
\sinh\left(\frac{t}
{\sqrt{k}}+t_0\right) \ .
\nonumber \\
\end{eqnarray}
We can fix the constant $t_0$ from the requirement
that at time $t=0$ the non-BPS Dp-brane 
sits at the point  $y=\pi r \  (\mT=0)$.
As a result   $t_0$ should be equal to zero. 
Then 
\begin{equation}
e=\frac{1}{\sqrt{1-\dot{\mT}^2_0}}
\Rightarrow
\dot{\mT}^2_0=\frac{\sqrt{e^2-1}}{e^2}
\end{equation}
and hence $e$ is related to the velocity
$\dot{\mT}$ at time $t=0$.

In summary, we have got the solution 
where 
 the spatial dependent 
tachyon condensation $T$ results  to
an emergence of  the array
of D(p-1)-branes and D(p-1)-antibranes.
Since non-BPS Dp-branes in type IIA (IIB)theories
have odd(even) $p$ the tachyon condensation leads
to the emergence of BPS D(p-1)-branes with 
even (odd) $p$. However the configuration 
when these  D-branes are inserted
in the background of $NS5$-branes is unstable and
hence 
these D-branes are moving towards to 
the  worldvolume of NS5-branes. This 
situation is described
by   time dependent condensation of  field $\mT$.

Let us now  consider  the  situation
when $\mT$ is function of $x$ and
 $T$ is function of $t$.  It is clear that 
we could proceed in the same way as
in the previous example however in order
to obtain clear physical meaning of the resulting
configuration it will be useful to construct
 the singular kink 
following the analysis performed in 
\cite{Sen:2003tm}. First of all, the equations
of motion  (\ref{eqmTT}) for
$\mT=\mT(x)$ and for $T=T(t)$  take the form 
\begin{eqnarray}\label{eqmTT2}
\frac{\sqrt{1-\dot{T}^2}}{\cosh
\frac{T}{\sqrt{2}}}
\left[-\frac{\sinh\frac{\mT}{\sqrt{k}}
\sqrt{1+\mT'^2}}
{\sqrt{k}\cosh^2\frac{\mT}{\sqrt{k}}}
-\partial_x\left(
\frac{1}{\cosh\frac{\mT}{\sqrt{k}}}
\frac{\partial_x\mT}
{\sqrt{1+\mT'^2}}\right)
\right]=0 \ , 
\nonumber \\
\frac{\sqrt{1+\mT'^2}}{\cosh
\frac{\mT}{\sqrt{k}}}
\left[-\frac{\sinh\frac{T}{\sqrt{2}}
\sqrt{1-\dot{T}^2}}
{\sqrt{2}\cosh^2\frac{T}{\sqrt{2}}}
+\partial_0\left(
\frac{1}{\cosh\frac{T}{\sqrt{2}}}
\frac{\partial_0T}
{\sqrt{1-\dot{T}^2}}\right)
\right]=0  \ .\nonumber \\
\end{eqnarray}
Now  the equation of motion for $\mT$ implies
\begin{equation}
\partial_x\left(
\frac{V(\mT,k)}{\sqrt{1+\mT'^2}}\right)=0 \ 
\end{equation}
that means that  the expression in the bracket
does not depend on $x$. Since 
for a kink solution $\mT\rightarrow \pm \infty$ 
 as $x\rightarrow \pm \infty$ and $V(\mT,k)
\rightarrow 0$ in this limit we obtain that 
the expression in the bracket vanishes for
$x\rightarrow \infty$ and from its independence
on $x$ it implies that it vanishes everywhere. 
This in turn implies that we should have
\begin{equation}
\mT=\pm \infty \ \mathrm{or}
\ \partial_x\mT=\infty \
\mathrm{(or \ both)} \   \mathrm{for \ all} \ x \ . 
\end{equation}
Clearly this solution looks singular. We will show,
following \cite{Sen:2003tm},
 that this solution has finite
energy density that is localised on codimension
one subspace however the interpretation is
slightly different than in the case of the tachyon 
kink on non-BPS Dp-brane in flat spacetime. 
 
To see this let us consider the field configuration
\begin{equation}\label{fd}
\mT(x)=f(ax) \ , 
f(u)=-f(-u) \ , f'(u)>0 \forall x \ , 
f(\pm \infty)=\pm \infty
\end{equation}
that in the limit $a\rightarrow \infty$
looks singular as expected.  
For this solution however we get
\begin{equation}
\frac{V(\mT,k)}
{\sqrt{1+\mT'^2}}=
\frac{V(f(ax),k)}{\sqrt{1+a^2f'^2(ax)}}
\end{equation}
that vanishes everywhere at 
the limit
$a\rightarrow \infty$ since the numerator
vanishes (except at $x=0$) and the
denominator blows up everywhere.  
Using this solution 
it is easy to find
other components of the stress energy tensor
\begin{eqnarray}
T_{00}(x)=\frac{\tau_pV(T,2)}{\sqrt{
1-\dot{T}^2}}V(\mT,k)
\sqrt{1+\mT'^2}=
\frac{\tau_pV(T,2)}{\sqrt{1-\dot{T}^2}}
V(f(ax),k)
af'(ax) \ , \nonumber \\
T_{ij}(x)=-\delta_{ij}\tau_pV(T,2)
\sqrt{1-\dot{T}^2}V(\mT,k)
\sqrt{1+\mT'^2}=\nonumber \\
=
-\delta_{ij}\tau_pV(T,2)
\sqrt{1-\dot{T}^2}
V(f(ax),k)
af'(ax)  
\nonumber \\
\end{eqnarray}
 in the limit $a\rightarrow \infty$.
Then the integrated $T_{00} \ ,
T_{ij}$ associated
with the codimension one solution are equal
to
\begin{eqnarray}
T^{kink}_{00}=
\int dx T_{00}=
\frac{\tau_pV(T,2)}{\sqrt{
1-\dot{T}^2}}\int dx V(f(ax),k)
af'(ax)=
\frac{\tau_pV(T,2)}{\sqrt{
1-\dot{T}^2}}\int dy V(y,k)dy \ ,
\nonumber \\
T^{kink}_{ij}=-\delta_{ij}
\tau_pV(T,2)
\sqrt{1-\dot{T}^2}
\int V(y)dy \ , 
\nonumber \\
\end{eqnarray}
where $y=f(ax)$. 
Thus 
$T^{kink}_{\alpha\beta} \ , \alpha , \beta=
0,2,\dots,p$ 
depend
on $V$ and not on the form of $f(u)$.
It is clear from the exponential
fall off the function of $V$ that
most of the contribution is contained
in the finite range of $y$. In fact,
in the limit $a\rightarrow \infty$
the stress energy tensor $T^{kink}_{\alpha\beta}$
is localised on codimension one
D(p-1)-brane with the tension given
as 
\begin{equation}
T_{p-1}=\tau_p\int dy V(y,k)=\tau_p
\int \frac{dy}{\cosh\frac{y}{\sqrt{k}}}
=\frac{(2\pi)\sqrt{k}}{2}\tau_p
=2\pi T_p^{non}R  
\end{equation}
using the fact that 
$\tau_p$ is equal to
\begin{equation}
\tau_p=\frac{2T^{non}_pR}
{\sqrt{k}} \ .
 \end{equation}
Finally we obtain
\begin{equation}
T_{00}^{kink}=\delta(x)
T_{p-1}\frac{V(T,2)}{\sqrt{1-\dot{T}^2}} \ ,
T_{ij}=-\delta(x)\delta_{ij}T_{p-1}V(T,2) 
\sqrt{1-\dot{T}^2} \ . 
\end{equation}
The geometrical meaning of this solution
is clear \cite{Kutasov:2004ct}. Since $\mT$ is directly 
related to the coordinate  $y$ that
parameterises the position of a non-BPS Dp-brane
on the transverse circle, 
the singular kink solution corresponds
to non-BPS D-brane that sits on top
of the fivebranes for all $x<0$  then at
$x=0$ goes around the y circle 
and then back to the fivebranes 
at $y=2\pi R$ 
where it stays for all $x>0$. 
This describes
non-BPS Dp-brane wrapped around the
transverse circle. Then the time dependent
solution of the equation of motion for $T$ 
\begin{equation}
\sinh \frac{T}{\sqrt{2}}=
\sinh\frac{t}{\sqrt{2}}
\end{equation}
describes the annihilation of this Dp-brane
to the closed string vacuum
\cite{Sen:2004nf}.

To obtain BPS-like D-brane that is stable
in the NS5-brane background we should
 consider the  situation when
both $\mT$ and $T$ are spatial 
dependent. For that reason
we  take following  ansatz
\begin{equation}\label{Tan}
T(x^1)=f_1(x^1) \ ,
x^1\equiv \xi^{p-1} \ ,
\mT(x^2)=f_2(x^2) \ ,
x^2\equiv \xi^p \ ,
\end{equation}
where $f_i(u)$ are functions 
with the properties given
in (\ref{fd}). For this ansatz
the matrix $\bI+\bM$ takes
the form
\begin{equation}
\bI+\bM=
\left(\begin{array}{ccc}
\bI_{(p-1)\times
(p-1)} & 0 & 0 \\
0 & 1+(\partial_1T)^2 & 0 \\
0 & 0 & 1+(\partial_2\mT)^2 \\ \end{array}
\right) \ ,
\end{equation}
where $\partial_1\equiv \partial_{x^1} \ ,
\partial_2\equiv \partial_{x^2}$.
Then the components of
the  stress energy tensor
are equal to
\begin{eqnarray}
T_{00}=\tau_pV(T,2)V(\mT,k)
\sqrt{(1+(\partial_2\mT)^2)
(1+(\partial_1T)^2)} \ , 
T_{0i}=0 \ , i=1,\dots, p  \ , 
\nonumber \\
T_{x^1x^1}=-\tau_pV(\mT,k)\sqrt{1+
(\partial_2\mT)^2}
\frac{V(T,2)}{\sqrt{1+(\partial_1T)^2}} \ ,
T_{x^1i}=0 \ , i=1,\dots,p-1 \nonumber \\
T_{x^2x^2}=\tau_pV(T,2)\sqrt{1+(\partial_1T)^2}
\frac{V(\mT,k)}{\sqrt{1+(\partial_2\mT)^2}} \ ,
T_{x^2i}=0 \ , i=1,\dots,p-2 \ , \nonumber \\
T_{ij}=-\delta_{ij}\tau_p
V(T,2)V(\mT,k)
\sqrt{(1+(\partial_2\mT)^2)
(1+(\partial_1T)^2)} \ , 
i,j=1,2,\dots,p-2 \ . 
\nonumber \\
\end{eqnarray}
Now the conservation of the
stress energy tensor implies
\begin{equation}
\partial_{x^1}T_{x^1x^1}=0 \ ,
\partial_{x^2}T_{x^2x^2}=0 \ .
\end{equation}
In other words,  $T_{x^1x^1}$ does not
depend on $x^1$ and 
$T_{x^2x^2}$ does not depend on
$x^2$. Since for $x^1\rightarrow \infty$
$V(T,2)\rightarrow 0$
and using the same arguments as
in the case given above 
  we get that $T_{x^1x^1}$ is equal to
zero  for 
all $x^1$. In
the same way one can argue that
 $T_{x^2x^2}=0$ for all $x^2$. 
Then the ansatz
(\ref{Tan}) has following physical interpretation:
 The
 condensation of the tachyon field
$T$ leads to the emergence of 
 BPS D(p-1)-brane localised at
the point $x^1=0$ on  the worldvolume of
non-BPS Dp-brane. Then the next
 condensation of the
field $\mT$ describes D(p-1)-brane that for
$x^2<0$ sits at the worldvolume of NS5-branes
at $y=0$  at the point $x^2=0$ wraps the
transverse circle back to $y=2\pi R$ and then
it sits on the worldvolume of NS5-branes
for $x^2>0$. In other words, this tachyon 
condensation leads to the BPS D(p-1)-brane
that wraps the transverse circle. As is well
known such a configuration is stable as
opposite to the case of the BPS D-brane 
whose worldvolume is parallel with the worldvolume
of the fivebranes.   

In order to  further support this picture we will 
 calculate the stress energy tensor
corresponding to this D(p-1)-brane 
\begin{eqnarray}
T_{\alpha\beta}^{D(p-1)}=
\int dx^1dx^2 T_{\alpha\beta}=
-\eta_{\alpha\beta}\tau_p\int dx^1 V(T,2)\sqrt{
1+T'^2}\int dx^2V(\mT,k)\sqrt{1+\mT'^2}=
\nonumber \\
=-\eta_{\alpha\beta}\tau_p\int V(y^1,2)dy^1
\int V(y^2,k)dy^2 \ , 
y^i=f_i(a_ix^i)  \ , i=1,2 \ ,
\nonumber \\
\end{eqnarray}
where $\alpha\ , \beta=0,1,\dots,p-2$. 
Since in the limit $a_i\rightarrow \infty$ the
tachyon potential vanishes almost everywhere 
except at the point $x^i=0$ we can write the
resulting stress energy tensor as
\begin{equation}
T_{\alpha\beta}=-\eta_{\alpha\beta}
\delta(x^1)\delta(x^2)T_{p-1} 
\end{equation}
where 
\begin{eqnarray}
T_{p-1}=\tau_p\int dy^1\frac{1}{\cosh\frac{y^1}{
\sqrt{2}}}\int dy^2\frac{1}{\cosh\frac{y^2}{\sqrt{k}}}
=\frac{T_{p-1}^{BPS}}{g_s}2\pi R
\nonumber \\
\end{eqnarray}
that is exactly an energy of BPS D(p-1)-brane
wrapped around the circle with radius $R$. 

Now we would like to determine the effective
action for translation zero modes of this
solution following the analysis performed in  
\cite{Sen:2003tm}. For that reason we will
consider the ansatz for  $T$
and $\mT$ 
\begin{equation}
T(x^1,\xi)=f_1(a_1(x^1-t^1(\xi)) \ ,
\mT(x^2,\xi)=f_2(a_2(x^2-t^2(\xi)) \ ,
\end{equation}
where we have denoted $\xi^{\alpha} \ ,
\alpha=0,1,\dots,p-2$ the coordinates
tangential to the kink worldvolume. 
For such a configuration we get
\begin{eqnarray}
\bA_{x^1x^1}=1+a_1^2f_1'^2 \ ,
\bA_{x^2x^2}=1+a_2^2f_2'^2 \ , 
\nonumber \\
\bA_{\alpha x^1}=\bA_{x^1\alpha}=
-a^2_1f'^2_1\partial_{\alpha}t^1 \ ,
\bA_{\alpha x^2}=\bA_{x^2\alpha}=
-a_2^2f'^2_2\partial_{\alpha}t^2 \ ,
\nonumber \\
\bA_{\alpha\beta}=(a_1^2f'^2_1-1)
\partial_{\alpha}t^1\partial_{\beta}t^1+
(a_2^2f'^2_2-1)
\partial_{\alpha}t^2\partial_{\beta}t^2+
\mathbf{a}_{\alpha\beta} \ , \nonumber \\
\mathbf{a}_{\alpha\beta}=
\eta_{\alpha\beta}+\partial_{\alpha}t^i
\partial_{\beta}t^i  \ .\nonumber \\
\end{eqnarray}
Let us  now define  following matrices
\begin{eqnarray}
\hat{\bA}_{\mu\beta}=
\bA_{\mu\beta}+
\bA_{\mu x^1}\partial_{\beta}t^1+
\bA_{\mu x^2}\partial_{\beta}t^2, \  
\hat{\bA}_{\mu x^1}=\bA_{\mu x^1}, \  
\hat{\bA}_{\mu x^2}=\bA_{\mu x^2} \ ,
\nonumber \\
\tilde{\bA}_{\alpha \nu}=
\hat{\bA}_{\alpha\nu}+
\hat{\bA}_{x^1\nu}
\partial_{\alpha}t^1+
\hat{\bA}_{x^2\nu}\partial_{\alpha}t^2, \ 
\tilde{\bA}_{x^1 
\nu}=\hat{\bA}_{x^1\nu}, \ 
\tilde{\bA}_{x^2 \nu}=\hat{\bA}_{x^2\nu} \ 
\nonumber \\
\end{eqnarray}
that obey 
\begin{equation}
\det\hat{\bA}=\det\tilde{\bA}=
\det \bA \ .
\end{equation}
On the other hand the explicit calculation 
gives  
\begin{eqnarray}
\tilde{\bA}_{\alpha\beta}=
\mathbf{a}_{\alpha
\beta} \ , 
\tilde{\bA}_{x^1\alpha}
=\tilde{\bA}_{\alpha x^1}=
\partial_{\alpha}t^1 \ ,
\tilde{\bA}_{x^2\alpha}=
\tilde{\bA}_{\alpha x^2}=
\partial_{\alpha}t^2 \ ,
\nonumber \\
\tilde{\bA}_{x^1x^1}=1+a^2_1f'^2_1 \ ,
\tilde{\bA}_{x^2x^2}=1+a^2_2f'^2_2 \ 
\nonumber \\
\end{eqnarray}
and hence the determinant $\det \tilde{\bA}$
for large $a_1, a_2$
takes the form
\begin{equation}
\det\tilde{\bA}=a^2_1f_1'^2a^2_2f'^2_2
\left[\det \mathbf{a}_{\alpha\beta}
+\mathcal{O}(\frac{1}{a_1^2})
+\mathcal{O}(\frac{1}{a_2^2})\right] \ . 
\end{equation}
Substituting this expression into the
effective 
action we get
\begin{eqnarray}
S=-\tau_p\int d^{p-1}\xi\sqrt{-\det 
\mathbf{a}}\int dx^1dx^2
 V(f_1(a_1),2)V(f_2
(a_2,k))a_1f'_1(a_1x^1)
a_2f'_2(a_2x^2)=\nonumber \\
=-\frac{T^{BPS}_{p-1}2\pi R}{g_s}
\int d^{p-1}\xi\sqrt{-\det 
\mathbf{a}} \nonumber \\
\end{eqnarray}
that is the right form of
the action for zero modes describing
transverse fluctuations of 
BPS D(p-1)-brane that wraps the
 circle with radius $R$.
\section{Conclusion}\label{fourth}
In this paper we have studied the
dynamics of a non-BPS Dp-brane 
in the background of $k$ $NS5$-branes
on transverse $R^3\times S^1$,
 following
   paper \cite{Kutasov:2004ct}, where
   the dynamics of  a 
BPS Dp-brane in this background 
was discussed.
 The main motivation
was to clarify the relation between 
the true tachyon mode on the worldvolume
of a non-BPS Dp-brane-that expresses an
instability of this object even in flat 
spacetime background-and the new 
tachyon field that arises from the redefinition
of the mode that describes the position 
of the non-BPS D-brane on a transverse circle $S^1$.
We have found that a non-BPS Dp-brane
in this background is invariant under 
the exchange $T$ with $\mT$ on condition
that  the numerical factors in the
 tachyon potentials are exchanged as well.
This
fact is clearly very puzzling since while 
$k$ is the number of $NS5$-branes and
hence has clear physical meaning 
the factor $2$ in the tachyon potential
$V$ does not have such a clear physical
 interpretation.
In the same way it is not clear whether 
the extended symmetry between $T,\mT$ and
parameters in the tachyon potentials that
was discussed in section (\ref{second}) has
some physical meaning or whether this is
only  pure coincidence. On the other
hand  
we mean that the idea that 
the tachyon on the worldvolume of
a non-BPS Dp-brane could have geometrical
origin is very intriguing and certainly 
deserve to be investigated further. 

Then in section (\ref{third})  we have  studied the solutions of the
tachyon effective action in the background
defined above. For the  spatial dependent
tachyon $T$ and time dependent tachyon
$\mT$ this solution describes collection of
 D(p-1)-branes and  D(p-1)-antibranes 
that move towards to the worldvolume of
fivebranes. On the other hand for the time dependent
$T$ and spatial dependent $\mT$ we have considered
the solution that
corresponds to the emergence of
a non-BPS D(p-1)-brane that 
wraps the transverse circle and that
 further annihilates
in the process of the time dependent 
tachyon condensation. 
An finally, we have constructed solutions 
where  both 
$T$ and $\mT$ were spatial dependent. 
We have then argued that this configuration
describes  BPS D(p-1)-brane wrapped 
around transverse circle $S^1$.   

We hope that this modest
contribution to the study of the dynamics of
BPS and non-BPS D-branes in the NS5-branes
background that we have performed in this
paper could be helpful for further research
of this very interesting subject and for 
better understanding of the role of the tachyon
in  string theory. 
\\
\\
{\bf Acknowledgement}

This work was supported by the
Czech Ministry of Education under Contract No.
14310006.


\begin{thebibliography}{20}


\bibitem{Sen:2004nf}
A.~Sen,
\emph{``Tachyon dynamics 
in open string theory,''}
arXiv:hep-th/0410103.

\bibitem{Sen:1999mg}
A.~Sen,
\emph{``Non-BPS states 
and branes in string theory,''}
arXiv:hep-th/9904207.

\bibitem{Witten:1998cd}
E.~Witten,
\emph{``D-branes and K-theory,''}
JHEP {\bf 9812} (1998) 019
[arXiv:hep-th/9810188].

\bibitem{Witten:2000cn}
E.~Witten,
\emph{``Overview of K-theory applied to strings,''}
Int.\ J.\ Mod.\ Phys.\ A {\bf 16} (2001) 693
[arXiv:hep-th/0007175].

\bibitem{Horava:1998jy}
P.~Horava,
\emph{``Type IIA D-branes, 
K-theory, and matrix theory,''}
Adv.\ Theor.\ Math.\ Phys.\  {\bf 2} (1999) 1373
[arXiv:hep-th/9812135].

\bibitem{Sen:1999md}
A.~Sen,
\emph{``Supersymmetric world-volume 
action for non-BPS D-branes,''}
JHEP {\bf 9910}, 008 (1999)
[arXiv:hep-th/9909062].

\bibitem{Garousi:2000tr}
M.~R.~Garousi,
\emph{``Tachyon couplings 
on non-BPS D-branes and 
Dirac-Born-Infeld action,''}
Nucl.\ Phys.\ B {\bf 584}, 284 (2000)
[arXiv:hep-th/0003122].

\bibitem{Bergshoeff:2000dq}
E.~A.~Bergshoeff, M.~de Roo, 
T.~C.~de Wit, E.~Eyras and S.~Panda,
\emph{``T-duality and 
actions for non-BPS D-branes,''}
JHEP {\bf 0005}, 009 (2000)
[arXiv:hep-th/0003221].

\bibitem{Kluson:2000iy}
J.~Kluson,
\emph{``Proposal for non-BPS D-brane action,''}
Phys.\ Rev.\ D {\bf 62}, 126003 (2000)
[arXiv:hep-th/0004106].

\bibitem{Lambert:2003zr}
N.~Lambert, H.~Liu and J.~Maldacena,
\emph{``Closed strings from decaying D-branes,''}
arXiv:hep-th/0303139.

\bibitem{Kutasov:2003er}
D.~Kutasov and V.~Niarchos,
\emph{``Tachyon effective 
actions in open string theory,''}
Nucl.\ Phys.\ B {\bf 666}, 56 (2003)
[arXiv:hep-th/0304045].

\bibitem{Niarchos:2004rw}
V.~Niarchos,
\emph{``Notes on tachyon 
effective actions and Veneziano amplitudes,''}
Phys.\ Rev.\ D {\bf 69}, 106009 (2004)
[arXiv:hep-th/0401066].

\bibitem{Kutasov:2004dj}
D.~Kutasov,
\emph{``D-brane dynamics near NS5-branes,''}
arXiv:hep-th/0405058.

\bibitem{Kutasov:2004ct}
D.~Kutasov,
\emph{``A geometric 
interpretation of the 
open string tachyon,''}
arXiv:hep-th/0408073.

\bibitem{Kluson:2004xc}
J.~Kluson,
\emph{``Non-BPS D-brane near NS5-branes,''}
arXiv:hep-th/0409298.

\bibitem{Saremi:2004yd}
O.~Saremi, L.~Kofman and A.~W.~Peet,
\emph{``Folding branes,''}
arXiv:hep-th/0409092.

\bibitem{Sahakyan:2004cq}
D.~A.~Sahakyan,
\emph{``Comments on D-brane 
dynamics near NS5-branes,''}
arXiv:hep-th/0408070.

\bibitem{Ghodsi:2004wn}
A.~Ghodsi and A.~E.~Mosaffa,
\emph{``D-brane dynamics in 
RR deformation of NS5-branes 
background and tachyon
cosmology,''}
arXiv:hep-th/0408015.

\bibitem{Panigrahi:2004qr}
K.~L.~Panigrahi,
\emph{``D-brane dynamics in Dp-brane background,''}
arXiv:hep-th/0407134.

\bibitem{Yavartanoo:2004wb}
H.~Yavartanoo,
\emph{``Cosmological solution 
from D-brane motion in 
NS5-branes background,''}
arXiv:hep-th/0407079.

\bibitem{Sen:2003tm}
A.~Sen,
\emph{``Dirac-Born-Infeld 
action on the tachyon kink and vortex,''}
Phys.\ Rev.\ D {\bf 68} (2003) 066008
[arXiv:hep-th/0303057].

\bibitem{Polchinski:1998rr}
J.~Polchinski,
\emph{``String theory. Vol. 2: 
Superstring theory and beyond,''}
\href{http://www.slac.stanford.edu/spires/find/hep/www?irn=4634802}
{SPIRES entry}


\bibitem{Kim:2003ma}
C.~Kim, Y.~Kim, O.~K.~Kwon and C.~O.~Lee,
\emph{``Tachyon kinks on unstable Dp-branes,''}
JHEP {\bf 0311} (2003) 034
[arXiv:hep-th/0305092].

\bibitem{Brax:2003rs}
P.~Brax, J.~Mourad and D.~A.~Steer,
\emph{``Tachyon kinks on non BPS D-branes,''}
Phys.\ Lett.\ B {\bf 575} (2003) 115
[arXiv:hep-th/0304197].

\bibitem{Kim:2003in}
C.~j.~Kim, Y.~b.~Kim and C.~O.~Lee,
\emph{``Tachyon kinks,''}
JHEP {\bf 0305} (2003) 020
[arXiv:hep-th/0304180].




\end{thebibliography}
\end{document}